\documentclass[prb, twocolumn,amsmath,amssymb, superscriptaddress]{revtex4-1}
\usepackage{graphicx}
\usepackage{natbib,hyperref}
\usepackage{amsmath,epsfig,color,amssymb}

\begin{document}

\author{Areg Ghazaryan}
\affiliation{Department of Physics, City College, City University of New York, New York, NY 10031, USA}
\author{Mohammad Hafezi}
\affiliation{Joint Quantum Institute, NIST and University of Maryland, College Park, MD 20742, USA}
\affiliation{Department of Electrical Engineering and IREAP, University of Maryland, College Park, MD 20742, USA}
\affiliation{Department of Physics, College Park, MD 20742, USA}
\author{Pouyan Ghaemi}
\affiliation{Department of Physics, City College, City University of New York, New York, NY 10031, USA}
\affiliation{Department of Physics, Grad. Center, City University of New York, New York, NY 10016, USA}
\title{Anisotropic exciton transport in transition-metal dichalcogenides}

\begin{abstract}
Due to the Coulomb interaction exciton eignestates in monolayer transitional metal dichalcogenides are coherent superposition of two valleys. 
The exciton band which couples to the transverse electric mode of light has parabolic dispersion for the center of mass momentum, whereas the one which couples to the transverse magnetic mode has both parabolic and linear components. In this work we present an experimental proposal to observe the signatures of linear component of the dispersion. In particular, it is demonstrated that by pumping the system with linearly polarized light the exciton transport is anisotropic compared to circularly polarized pump. We show that the results persist for moderate level of disorder present in realistic systems. Finally, we demonstrate that similar effects can be obtained for positively detuned exciton-polaritons, in less stringent experimental requirements compared to bare exciton case.
\end{abstract}

\maketitle

\section{Introduction}
\label{Intro}
Graphene, single layer of graphite, has been the subject of intense research for over a decade due to its special Dirac type bands arising from its honeycomb lattice structure and its true two-dimensional (2D) nature \cite{CastroNeto2009,Goerbig2011}. The main obstacle for utilizing graphene in optoelectronics is the gapless nature of its bands, which is protected by inversion symmetry. Therefore, numerous methods have been proposed to brake inversion symmetry of graphene and open a gap \cite{McCann2013}. Nevertheless, the gap obtained with this methods is usually small and not suitable for near-infrared or visible light experiments. Observation of a direct band gap in monolayer transition metal dichalcogenides (TMDs) \cite{Mak2010,Splendiani2010,Duan2015}, such as $\mathrm{MoS}_2,\mathrm{MoSe}_2,\mathrm{WS}_2,\mathrm{WSe}_2$, with the size of the order of visible light frequency opened new avenues in this regard. In TMDs, transition metal atom is coordinated with six chalcogen atoms in trigonal prismatic geometry. The lattice structure of TMDs is similar to graphene, but the inversion symmetry is explicitly broken, because of alternating metal and chalcogen atoms. It possesses a gap in the range of $1.4-1.7\,\mathrm{eV}$ \cite{Xiao2012} and due to the almost 2D nature strongly bonded excitons\cite{Berkelbach2013,He2014,Zhang2014,Berkelbach2015}. The binding energy reaches up to $500\,\mathrm{meV}$, therefore excitonic related phenomena can be realized at room temperatures in these materials. Similar to graphene, the low energy band structure of TMDs consists of two degenerate valleys at the corners of the Brillouin zone (BZ). Since Berry curvature has opposite sign for two valleys only one circular polarization of the light couples to the one specific valley \cite{Xiao2012}. This makes TMDs attractive for valleytronic applications. TMDs possess strong spin-orbit coupling (SOC) originating from transition metal $d$ bands, which polarizes the spin direction at the top of the valence band in each valley \cite{Xiao2012}. There is also spin splitting due to the SOC in the conduction band, although it is of magnitude smaller compared to valence band \cite{Kormanyos2015}. Due to this splittings the constituent particles of the bright low energy excitons at each valley are characterized with specific spin, which is opposite between the valleys. These correlation between valley and spin indices can be utilized for realizing both valley and spin coherence, which was demonstrated experimentally both in low and room temperatures \cite{Zeng2012,Mak2012,Cao2012,Wang2016}.

Exciton-polariton (EP) \cite{Savona1999,Carusotto2013,Deng2010} states are also extensively studied in TMDs due to the large light-matter coupling originating from tightly bound excitons. EPs are hybrid quasiparticles comprised of exciton and photon, which are realized when the system is placed at the antinode of the planar optical cavity. Compared to excitons, EPs have large coherence length, due to the spatial spreading of the photon and are therefore, less susceptible to disorder scattering \cite{Savona2007}. More importantly, because the masses of the exciton and photon are very different by changing the cavity detuning, the effective mass of polariton can be largely  modified. 
In addition, the exciton part of polariton results in an enhanced $\chi^{(3)}$ (third component of the electric susceptibility) nonlinearity, with considerable polariton-polariton interaction. These peculiar characteristics have found their realization in numerous physical phenomena, such as optical spin Hall effect \cite{Kavokin2005,Leyder2007,Shelykh2009} and Bose-Einstein condensation of polariton liquid \cite{Kasprzak2006,Balili2007,Sun2017}. Due to the strong light-matter coupling in TMDs EPs have been successfully realized in strong coupling regime both in low and room temperatures \cite{Liu2015,Dufferwiel2015,Wang2016Coh}. More robust valley coherence of EPs compared to bare excitons in TMDs has been also demonstrated \cite{Sun2017VP,Dufferwiel2017,Chen2017}.

\begin{figure}
\includegraphics[width=8.5cm]{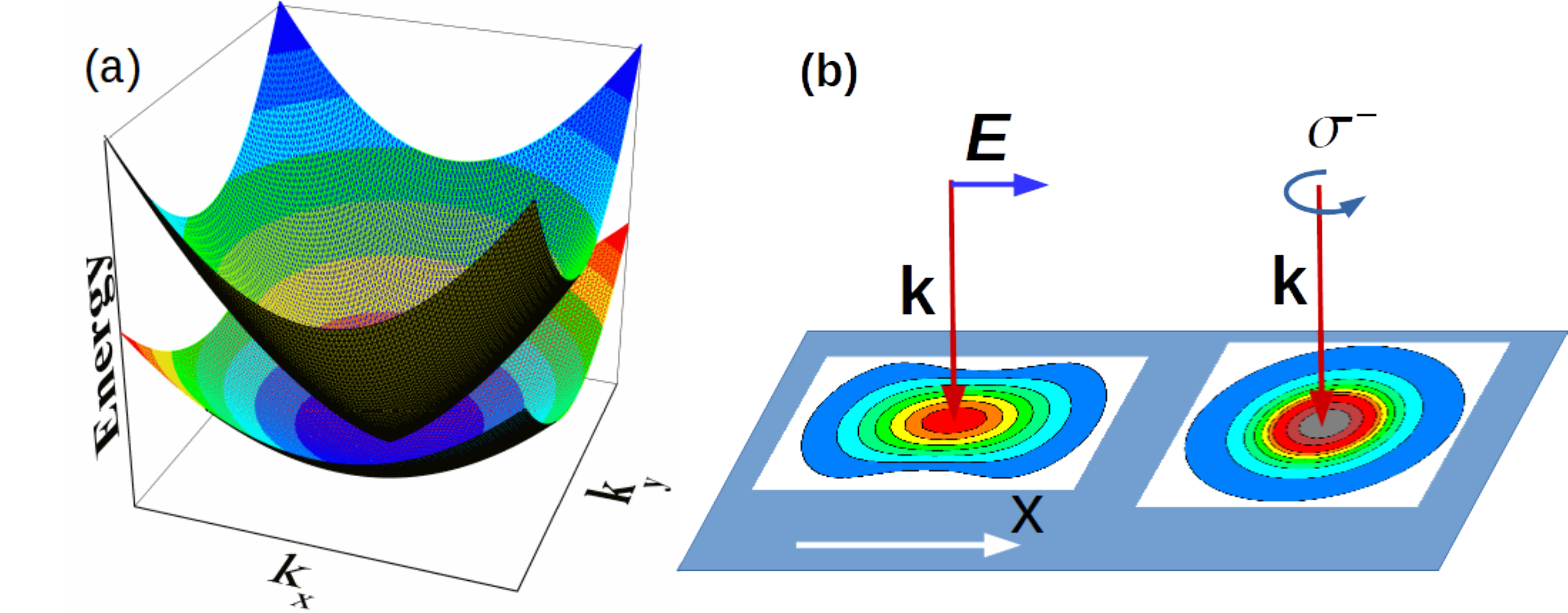}
\caption{\label{fig:SchemeFig} (a) Schematic drawing of the band structure of excitons in TMDs and (b) depiction of exciton transport when the system is pumped with linearly polarized light (left) and right circularly polarized light (right).  
}
\end{figure}

A distinctive feature of excitons in TMDs, in comparison with III-V and II-VI semiconductor systems (such as GaAs, CdTe etc.), is in the form of their dispersion as a function center of mass (CM) momentum. It is now well established that due to the exchange interaction the exciton bound states are superposition of excitonic states in two valleys \cite{Yu2014,Qiu2015,Wu2015,Yu2014PRB,Gartstein2015}. The lowest energy exciton is described by a parabolic dispersion, whereas the higher one has both parabolic and linear components (Fig.~\ref{fig:SchemeFig} (a)) \cite{Qiu2015,Wu2015,Yu2014PRB,Gartstein2015}. These two bands couple to the transverse electric (TE) and transverse magnetic (TM) components of light, respectively. This novel excitonic band structure is not experimentally verified yet. The direct measurement of the exciton dispersion using EPs is hindered, due to the fact that in currently accessible samples, for momenta accessible in EP measurement, the energy gap between two exciton bands is comparable to the  exciton linewidth. In addition, even though EP transport is less sensitive to the disorder in the sample, the steep photon's dispersion makes it harder to observe the signature of linear component of excitonic bands in EP diffusion measurements. In this work we propose a method to observe the signatures of the novel form of the exciton dispersion through anisotropic exciton transport. Particularly, we demonstrate that when the system is excited with linearly polarized light, the non-polarized emission is anisotropic, whereas it becomes isotropic for circularly polarized pump (Fig.~\ref{fig:SchemeFig} (b)). This anisotropic transport is robust against moderate levels of disorder potential. We also analyze the transport phenomenon for EPs and show that similar anisotropic effect can be observed also for considerably positively detuned EPs. For GaAs type semiconductors both exciton bands are parabolic and exciton transport is isotropic regardless of the polarization of the pump when similar setup is used as for TMDs. 
There were several attempts to experimentally probe exciton transport in TMDs using transient absorption microscopy \cite{Cui2014,Yuan2017}, direct charge-coupled device imaging method \cite{Kato2016}, spatially-resolved steady state and time-resolved photo-luminescence \cite{Mouri2014,Cadiz2018}. Obtained experimental results were analyzed using 2D diffusion equation\cite{Kato2016,Yuan2017}, which gives consistent results with the experiments for non polarization resolved transport dynamics. In these experiments on TMDs both excitation and detection are not polarization resolved and also the system is pumped with much larger energy compared to exciton resonance. Polarization resolved transport of EPs with polarization beats appearing at specific directions has been experimentally demonstrated for AlGaAs quantum well with pulsed laser pump \cite{Langbein2007}. Our results show that such measurements on TMDs can reveal much more information about the strong particle-hole interactions which lead to the formation of excitons. This also points to the importance of improvement in quality of TMD samples.

The paper is organized as follows. In Sec.~\ref{Model} we present effective Hamiltonians for calculating exciton and EP dispersions. We also discuss time dependent Schr\"odinger equations resulting from effective Hamiltonians and the numerical procedure for simulating the transport dynamics of the excitons and EPs. In Sec.~\ref{Results} we present obtained results derived from the framework outlined in Sec.~\ref{Model} and discuss the potential application of the system in optical switch settings. Finally, we summarize and present some concluding remarks in Sec.~\ref{Conclusions}. 


\section{The Model}
\label{Model}

The two dimensional structure of TMDs leads to weak screening of Coulomb interaction and very large excitonic binding energy. Theoretical studies of excitonic spectrum in TMDs have shown that the interplay of their electronic band structure and Coulomb interaction leads to a distinctive dispersion relation for excitons in TMDs \cite{Qiu2015,Wu2015}. This excitonic band structure, for small center of mass momenta, can be expressed using the following effective Hamiltonian:

\begin{equation}
\mathcal{H}_\mathrm{ex}(\mathbf{k})=\left(\begin{array}{cc}
h_\mathrm{ex}(k) & S^{-}_\mathrm{ex}(\mathbf{k}) \\
S^{+}_\mathrm{ex}(\mathbf{k}) & h_\mathrm{ex}(k)
\end{array}\right),
\label{ExcHam}
\end{equation}                
where $\mathbf{k}=ke^{i\varphi}$, $h_\mathrm{ex}(k)=E^{0}_\mathrm{ex}+\frac{\hbar^2k^2}{2m_\mathrm{ex}}+\alpha k$, $S^{\pm}_\mathrm{ex}(\mathbf{k})=\alpha k e^{\pm i2\varphi}$. Here $E^{0}_\mathrm{ex}$ corresponds to the energy of the exciton at $\mathbf{k}=0$ measured from the top of the valence band and this is the energy observed in photoluminescence measurements. In the calculation we use the values $m_\mathrm{ex}=0.6m_0$ ($m_0$ being free electrons mass) for exciton effective mass and $\alpha=90\,\mathrm{meV}\cdot\mathrm{nm}$ for SOC amplitude. The two spinor components of the Hamiltonian (\ref{ExcHam}) correspond to the two valleys or spins of the constituent electron-hole of exciton. As was noted above optical transitions in each valley couple to specific circular polarization of light, therefore, equally Hamiltonian (\ref{ExcHam}) is written in circular polarization basis. It has two eigenvalues, the lowest in energy having parabolic dispersion $E^{1}_\mathrm{ex}(\mathbf{k})=E^{0}_\mathrm{ex}+\frac{\hbar^2k^2}{2m_\mathrm{ex}}$, while the higher one has both linear and parabolic components $E^{2}_\mathrm{ex}(\mathbf{k})=E^{0}_\mathrm{ex}+2\alpha k+\frac{\hbar^2k^2}{2m_\mathrm{ex}}$. It can be shown that lower band couples to the TE and upper one to TM linear polarized light \cite{Qiu2015}.

The Hamiltonian describing EPs can be written as exciton-photon coupled system \cite{Bleu2017}
\begin{equation}
\mathcal{H}_\mathrm{ep}(\mathbf{k})=\left(\begin{array}{cccc}
h_\mathrm{ex}(k) & S^{-}_\mathrm{ex}(\mathbf{k}) & V & 0 \\
S^{+}_\mathrm{ex}(\mathbf{k}) & h_\mathrm{ex}(k) & 0 & V \\
V & 0 & h_\mathrm{p}(k) &  S^{-}_\mathrm{p}(\mathbf{k}) \\
0 & V & S^{+}_\mathrm{p}(\mathbf{k}) & h_\mathrm{p}(k)
\end{array}\right),
\label{EPHam}
\end{equation}
where photon is describe by a parabolic dispersion $h_\mathrm{p}(k)=E^{0}_\mathrm{p}+\frac{\hbar^2k^2}{2m_\mathrm{p}}$ with effective mass $m_\mathrm{p}=4\times10^{-5}m_0$. The non-diagonal term in photon part of the Hamiltonian $S^{\pm}_\mathrm{p}(\mathbf{k})=\beta k^2 e^{\pm i2\varphi}$ (with double winding and amplitude $\beta$) is due to the TE-TM splitting \cite{Panzarini1999} and we use the value $\beta=47300\,\mathrm{meV}\cdot\mathrm{nm}^2$. Coupling amplitude between exciton and photon parts of Hamiltonian (\ref{EPHam}) is related to Rabi frequency $\Omega$ by $V=\hbar\Omega/2$ and $\hbar\Omega=20\,\mathrm{meV}$. The detuning of EP system is defined as $\Delta=E^{0}_\mathrm{p}-E^{0}_\mathrm{ex}$. The Hamiltonian (\ref{EPHam}) has four eigenvalues, two of which inherit linear component of dispersion from excitonic component, although observation of that linear component for EPs depends strongly on detuning as will be discussed below.

\begin{figure}
\includegraphics[width=8.5cm]{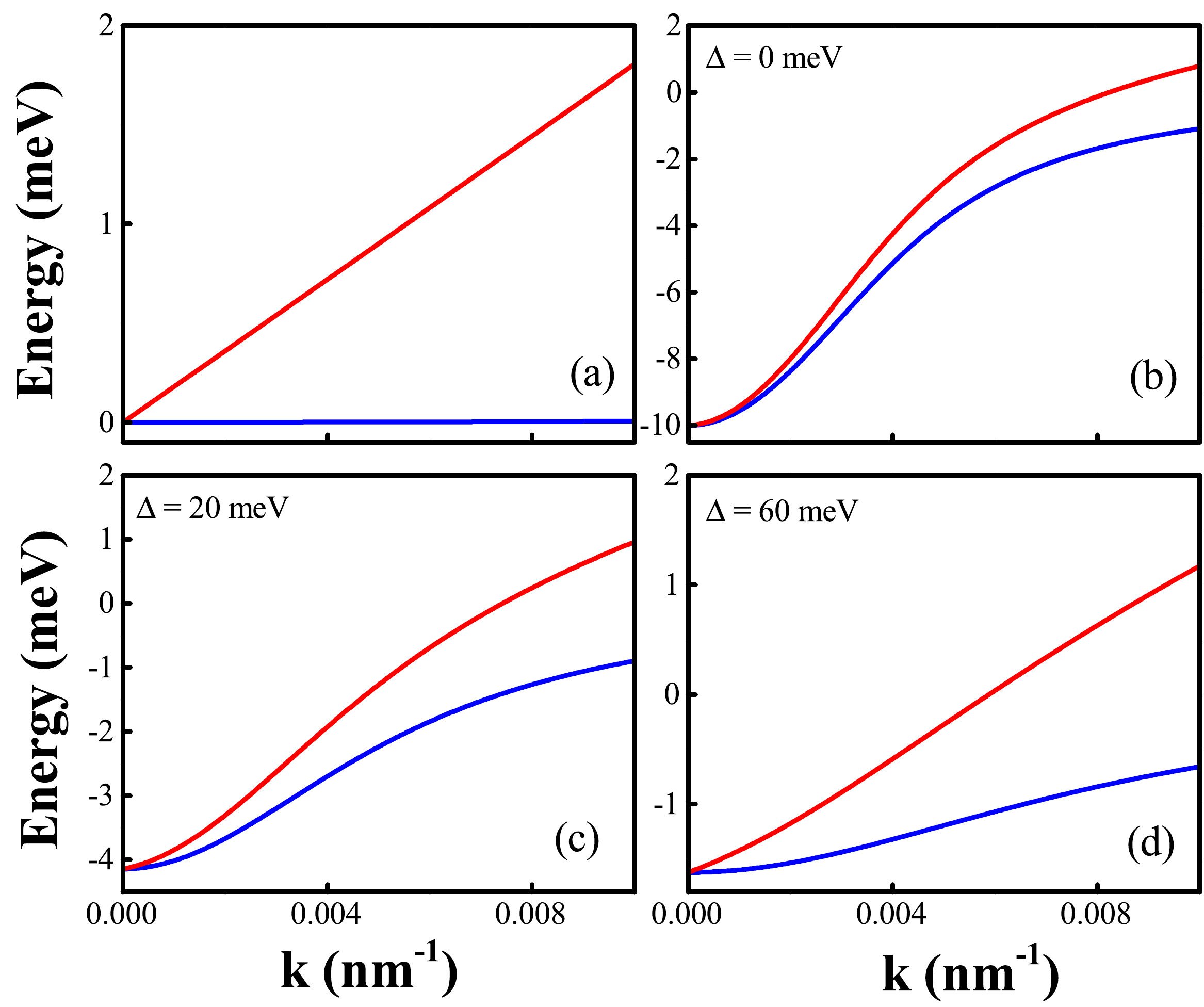}
\caption{\label{fig:EnDepk} The dependence of two exciton branches resulting from (\ref{ExcHam}) (a) and lower two EP branches from (\ref{EPHam}) for detunings $\Delta=0\,\mathrm{meV}$ (b), $\Delta=20\,\mathrm{meV}$ (c), $\Delta=60\,\mathrm{meV}$ (d) on absolute value of in plane momentum $k$.  
}
\end{figure}

In order to calculate spatial dynamics of excitons we numerically solve time-dependent Schr\"odinger equation based on Hamiltonian (\ref{ExcHam}), namely
\begin{align}
i\hbar\frac{\partial \phi_R(\mathbf{r},t)}{\partial t}&=h_\mathrm{ex}(k)\phi_R(\mathbf{r},t)+S^{-}_\mathrm{ex}(\mathbf{k})\phi_L(\mathbf{r},t)- \nonumber \\
&\frac{i\hbar}{2\tau_{ex}}\phi_R(\mathbf{r},t)+U(\mathbf{r})\phi_R(\mathbf{r},t)+P_R(\mathbf{r}), \nonumber \\
i\hbar\frac{\partial \phi_L(\mathbf{r},t)}{\partial t}&=h_\mathrm{ex}(k)\phi_L(\mathbf{r},t)+S^{+}_\mathrm{ex}(\mathbf{k})\phi_R(\mathbf{r},t)- \nonumber \\
&\frac{i\hbar}{2\tau_{ex}}\phi_L(\mathbf{r},t)+U(\mathbf{r})\phi_L(\mathbf{r},t)+P_L(\mathbf{r}), 
\label{ExcSch}
\end{align}
where $\phi_R(\mathbf{r},t)$ and $\phi_L(\mathbf{r},t)$ correspond to exciton wave function in two valleys (or to the right and left circular polarization, respectively). The double winding of $S^{\pm}_\mathrm{ex}(\mathbf{k})$ accounts for orbital momentum difference between right and left circularly polarized light. $\tau_{ex}=5.3\,\mathrm{ps}$ is the exciton lifetime, $U(\mathbf{r})$ is the disorder potential acting on the excitons and $P_R(\mathbf{r})$ and $P_L(\mathbf{r})$ are the continuous-wave (cw) pump terms for right and left circular polarization respectively. The exciton disorder potential is assumed Gauss correlated in space \cite{Savona2006,Savona2007,Sarchi2009}
\begin{equation}
\langle U(\mathbf{r})U(\mathbf{r^\prime})\rangle=\sigma^2e^{-|\mathbf{r}-\mathbf{r}^\prime|^2/\xi^2},
\end{equation}
where we take correlation length $\xi=10\,\mathrm{nm}$ and $\sigma$ is the correlation amplitude. After discretizing the space for numerical integration, the disorder potential can be calculated by the formula 
\begin{equation}
U(\mathbf{r})=\frac{2\sigma\sqrt{\Delta x\Delta y}}{\xi\sqrt{\pi}}\sum_{\mathbf{r}^\prime}c(\mathbf{r}^\prime)e^{-2|\mathbf{r}-\mathbf{r}^\prime|^2/\xi^2},
\end{equation}
where $\langle c(\mathbf{r})c(\mathbf{r}^\prime)\rangle=\delta_{\mathbf{r}\mathbf{r}^\prime}$ and $\Delta x$, $\Delta y$ denote discretization steps in $x$ and $y$ directions respectively. 
We model the pump by the Gaussian shape in momentum space \cite{Leyder2007,Liew2009}
\begin{equation}
P_{R,L}(\mathbf{k},t)=P^0_{R,L}e^{-\left(\mathbf{k}-\mathbf{k}_p\right)^2L^2/4}\frac{i\Gamma e^{-iE_pt/\hbar}}{h_{ex}(k)-E_p-i\Gamma},
\label{pump}
\end{equation}
where $\mathbf{k}_p$ is the pump momentum, $E_p$ is the pump energy, $\Gamma$ is the linewidth, $L$ pump spot size. We take pump energy $E_p$ to be the zero energy of the system in order to make pump term time independent.

As for the excitons similar time dependent Schr\"odiner equations can be written for polaritons:
\begin{align}
i\hbar\frac{\partial \psi(\mathbf{r},t)}{\partial t}&=h_\mathrm{p}(k)\psi(\mathbf{r},t)+S_\mathrm{p}(\mathbf{k})\psi(\mathbf{r},t)-\nonumber \\
&\frac{i\hbar}{2\tau_{p}}\psi(\mathbf{r},t)+V\phi(\mathbf{r},t)+P(\mathbf{r}), \nonumber\\
i\hbar\frac{\partial \phi(\mathbf{r},t)}{\partial t}&=h_\mathrm{ex}(k)\phi(\mathbf{r},t)+S_\mathrm{ex}(\mathbf{k})\phi(\mathbf{r},t)- \nonumber \\
&\frac{i\hbar}{2\tau_{ex}}\phi(\mathbf{r},t)+V\psi(\mathbf{r},t)+U(\mathbf{r})\phi(\mathbf{r},t),
\label{EpSch}
\end{align}
where now $\psi(\mathbf{r},t)$ and $\phi(\mathbf{r},t)$ are two component spinors (denoting the two valleys or two circular polarizations) corresponding to wave function for photon and exciton respectively. Similar to $\tau_{ex}$, $\tau_{p}=1.3\,\mathrm{ps}$ is the photon lifetime in the cavity. In (\ref{EpSch}) the disorder potential is only acting on exciton component of polariton and pump is generating photons, which get converted into polaritons through exciton-photon coupling. In this case pump has also two components which determine the polarization of the pump. Each component of the pump can be calculated by (\ref{pump}) by only changing the $h_{ex}(k)$ in the denominator with corresponding polariton dispersion.

We numerically solve (\ref{ExcSch}) and (\ref{EpSch}) on a $256\times256$ grid in real space of the size $3\mathrm{\mu m}\times3\mathrm{\mu m}$. For the terms in Hamiltonian which are diagonal in momentum space we do Fourier transform of the wave function from real space, calculate the action of those terms and then transform back \cite{Bleu2017} using fast Fourier transform algorithm \cite{PressBook}. The time step for integration of the equations (\ref{ExcSch}) and (\ref{EpSch}) is $\Delta t=10^{-4}\,\mathrm{ps}$ and we run the simulation until the populations of the particles stabilize (around $40\,\mathrm{ps}$). For integration we use Adams-Bashforth-Moulton method \cite{PressBook} in nVidia CUDA GPU framework.

\begin{figure}
\includegraphics[width=8cm]{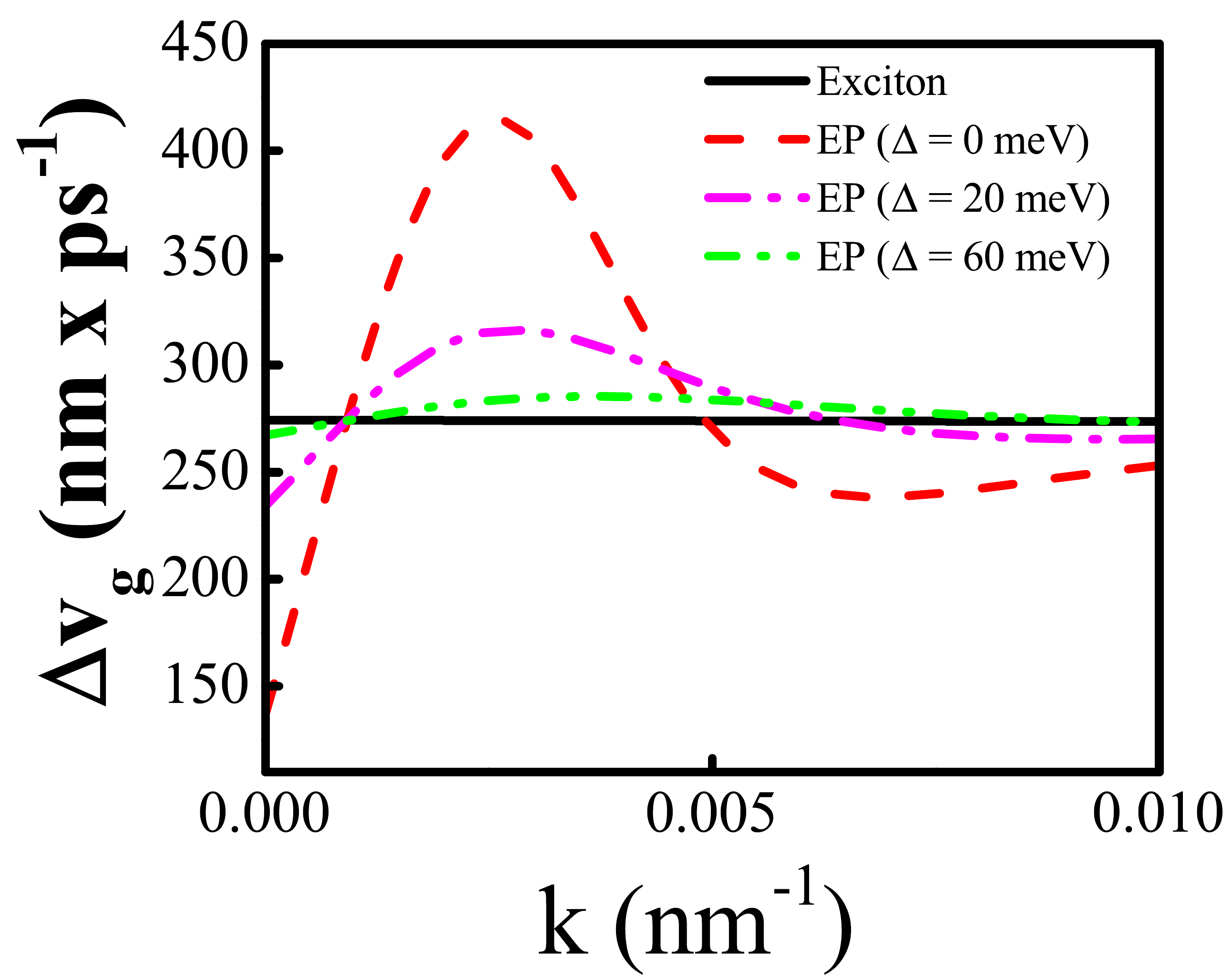}
\caption{\label{fig:DVDepk} The dependence of the difference of group velocities of two branches of exciton and EP shown in Fig.~\ref{fig:EnDepk} on absolute value of in plane momentum $k$.  
}
\end{figure}

\section{Results and Discussion}
\label{Results}
As was noted above one of the special characteristics of the excitons in TMDs is the presence of the linear component in the dispersion of one of the excitonic branch. Due to the fact, that EP is a superposition of exciton and photon, it inherits this linear component of dispersion. Fig.~\ref{fig:EnDepk} shows the dependence of the two excitonic and two lower EP branches (for detunings $\Delta=0,20,60\,\mathrm{meV}$) on absolute value of in plane momentum $k$. The momentum range presented in the figures correspond to the range accessible in angle resolve absorption and photoluminescence experiments on TMDs. As was noted above and can be seen from Fig.~\ref{fig:EnDepk} (a) one of the branches has linear dependence on momentum. Besides, both exciton energies have also parabolic term, which is extremely small for the values of momenta considered and the lowest exciton branch appears to be flat. The main objective of the current work is to find experimental signature which will unambiguously demonstrate the presence of the linear component in exciton dispersion. In Fig.~\ref{fig:EnDepk} (b-d) we also present two lower EP dispersion for different detunings. While linear component is present in the upper energy branch for all values of detuning, it becomes pronounced only for large positive detunings ($\Delta=60\,\mathrm{meV}$). This is already an indication, that using EPs in the current setup is not going to be as beneficial as is the case for other setups involving EP transport (such as optical spin Hall effect\cite{Kavokin2005,Leyder2007}), because the feature which we are interested shows up only for EPs which are comprised mostly from exciton component. This is related to the fact, that photon has extremely small effective mass compared to exciton and although exciton has linear component in its dispersion, for negative or close to zero detuning photon component will prevail already at extremely small momenta, making the experimental observation of linear component unrealistic due to the finite momentum broadening of the pump.

The essential difference between linear and parabolic bands which serves as a motivation for the experimental setup presented here is the difference between group velocities of two branches close to $k=0$ momentum. In Fig.~\ref{fig:DVDepk} the dependence  of the difference of group velocities ($v_g=\frac{1}{\hbar}\frac{\partial E(k)}{\partial k}$) of two branches shown in Fig.~\ref{fig:EnDepk} is presented on absolute value of in plane momentum $k$. For the case of excitons the difference is constant $\Delta v_g=2\alpha/\hbar\approx274.3\,\mathrm{nm}/\mathrm{ps}$. As was evident already from Fig.\ref{fig:EnDepk} for EPs the group velocity difference gets decreased for small momenta compared to exctions. The peak at $k=3\,\mathrm{\mu m}^{-1}$ observed for the group velocity difference of EPs is due to the TE-TM splitting of cavity, which is parabolic. With the increase of the value of $k$ the EP becomes more exciton like and the group velocity difference converges to the constant value for excitons.

\begin{figure}
\includegraphics[width=8.5cm]{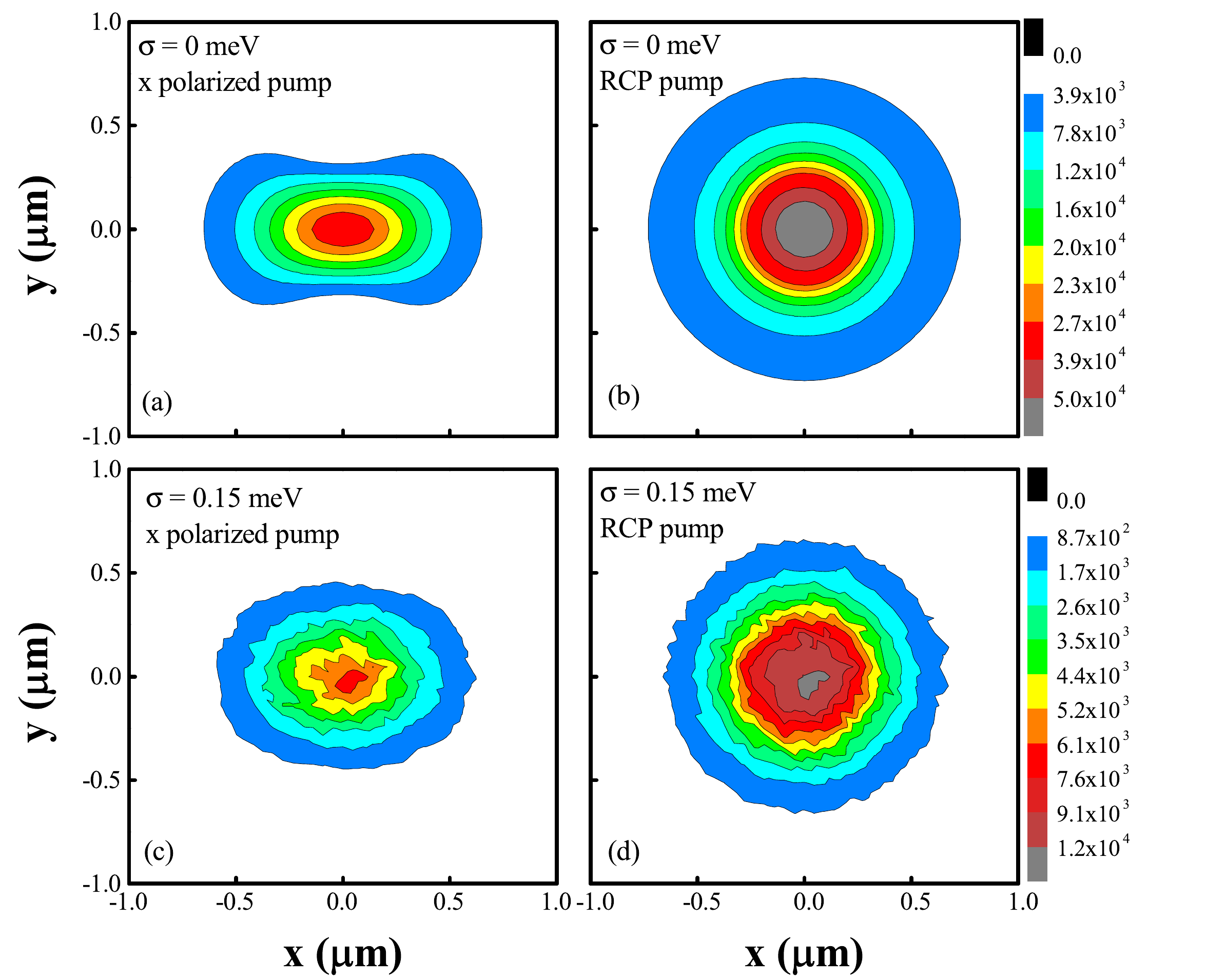}
\caption{\label{fig:ExcTrans} Exciton transport when pumped with $x$ polarized pump (a,c) and right circularly polarized (RCP) pump (b,d). (a,b) correspond to the case without disorder, (c,d) for the case with disorder with correlation amplitude $\sigma=0.15\,\mathrm{meV}$.   
}
\end{figure}

In order to study large group velocity difference at small momenta due to the linear component of the dispersion we propose exciton or EP anisotropic transport experiment using polarized pump similar to the experiment performed for AlGaAs quantum well\cite{Langbein2007}, although for current setup polarization resolved detection is not essential. The fact that excitons have finite lifetime and group velocity difference due to the TE-TM splitting of the cavity prevails at intermediate momenta for EPs the form and energy of the pump plays essential role in current setup. It is clear that pump should operate close to the minimum of two branches of excitons or lower EPs. Besides that for EPs it is desirable that pump spot size to be large enough, so that momentum broadening of the pump does not overlap with the TE-TM splitting dominated region of group velocity difference. Finally, the finite lifetime of exciton or EP poses upper bound on pump spot size, so that transport anisotropy region manages to escape pumping region and is detectable in the experiment. Therefore, we use the pump with parameters $E_p=E^0_\mathrm{ex}$ (or equal to the lowest solution of $\mathcal{H}_\mathrm{ep}(\mathbf{0})$ for EPs), $\mathbf{k}_p=0$, $\Gamma=0.2\,\mathrm{meV}$, $L=400\,\mathrm{nm}$, $P^0=10^4$.

In Fig.~\ref{fig:ExcTrans} results of exciton transport are presented when the system is pumped with $x$ linearly ($x$ denoting the direction of linear pump polarization) and right circularly polarized (RCP) pumps. The emitted light is not polarization resolved. As can be seen from Fig.~\ref{fig:ExcTrans} (a) for $x$ polarized pump the obtained signal is anisotropic. As was mentioned above the fully parabolic branch of exciton couples to TE polarization, whereas the branch which has both linear and parabolic components couples to TM polarization of light. Therefore, when the system is pumped with $x$ polarized light it couples to upper branch of excitons in $x$ direction and lower branch in $y$ direction. Due to the fact that upper branch has larger group velocity compared to lower branch this results in observed anisotropy. At intermediate directions $x$ polarized light is not an eigenstate of the system and therefore, polarization direction is rotated as excitons move away from pumping spot due to the SOC of excitons. The largest rotation happens at diagonal directions \cite{Langbein2007}. Therefore, at intermediate angles the distance of propagation depends on which proportion of its lifetime exciton is in the eigenstate of TM polarization. For the case of RCP pump shown in Fig.~\ref{fig:ExcTrans} (b), the signal is symmetric, because for all directions RCP light can be decomposed into TE and TM components with equal amplitude, so TM component is always present and dominates the propagation. It should be noted, that we have considered other types of dispersions for excitons and anisotropic propagation is only present for this special case. In particular, in the initial work on TMD excitons \cite{Yu2014}, it was argued that the two excitons can be described with linear dispersions of Dirac massless particles. While it was later clarified, that this calculation does not include intravalley exchange interaction \cite{Qiu2015,Wu2015,Gartstein2015}, our current setup gives the possibility to experimentally probe the actual dispersion of excitons. When considering either linear dispersions of Dirac massless particles or both parabolic dispersions (as in III-IV or II-VI semiconductors) for two exciton branches the signal is always isotropic when the pump parameters described above has been used. As is shown in Fig.~\ref{fig:ExcTrans} (c,d) the anisotropy is present also for the case of disorder with correlation amplitude $\sigma=0.15\,\mathrm{meV}$. In current TMD samples disorder is quite strong and the transport is dominated by the disorder, which makes the observation of the presented anisotropic transport cumbersome. Despite that, recent advances in sample preparation and exciton diffusion measurement \cite{Cadiz2018} make the transport signatures presented in the current paper within experimental reach in upcoming years.

\begin{figure}
\includegraphics[width=8.5cm]{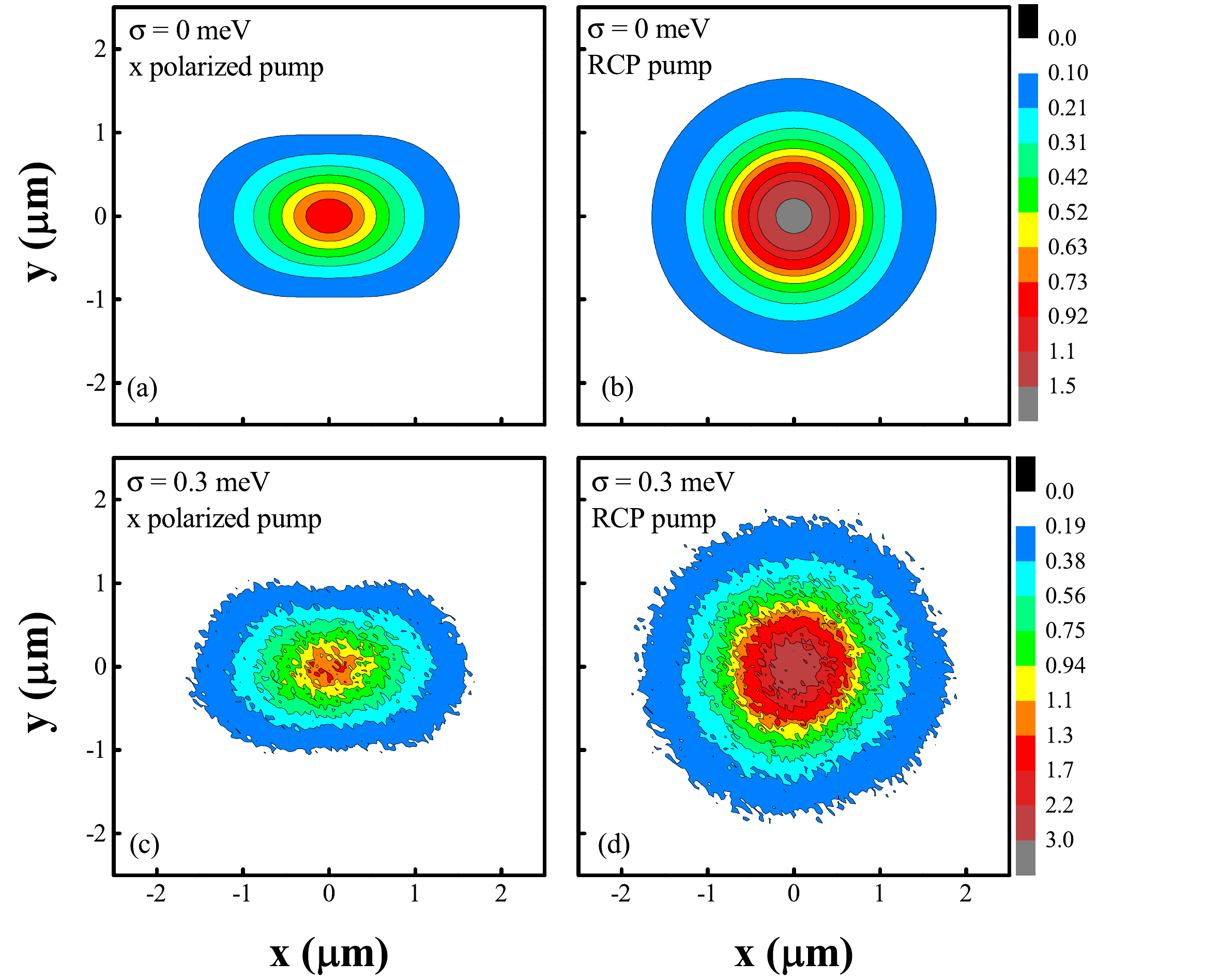}
\caption{\label{fig:EPTrans} EP transport when pumped with $x$ polarized pump (a,c) and right circularly polarized (RCP) pump (b,d). (a,b) correspond to the case without disorder, (c,d) for the case with disorder with correlation amplitude $\sigma=0.3\,\mathrm{meV}$. The detuning of EPs in the calculations is $\Delta=60\,\mathrm{meV}$.   
}
\end{figure}

The anisotropic transport setup can also be using in optical switch settings. In particular consider initial RCP pump. As was noted above the propagation is isotropic in this case. Now consider second pump at the same spot. If the second pump is again RCP and has the same phase as initial one, the signal will not change. Whereas, if the second pump is left circularly polarized (LCP), then sum of two  pumps will be $x$ linearly polarized and the signal will be anisotropic, which could be detected perpendicular to the overall pump polarization direction.

Finally, we discuss whether this effect can be observed for EPs. As was noted above the group velocity difference close to $k=0$ drops down considerably with the decrease of detuning. Comparable group velocity difference to excitons can be obtained for the detuning $\Delta=60\,\mathrm{meV}$. In this case at $k=0$ the proportion of exciton component of EP is 97\%, so the impact of photon component should not be substantial. In Fig.~\ref{fig:EPTrans} the results of EP transport are presented when the system is pumped with $x$ polarized and RCP pumps as in Fig.~\ref{fig:ExcTrans}. As can be seen from Fig.~\ref{fig:EPTrans} (a) anisotropic transport is observed for this case as well. Due to the photonic component of EPs, the overall group velocities have increased, so that the signal propagates at much longer distances than with excitons. This makes the observation of the signal less troublesome, because of the clear separation of the signal and pump regions. In terms of disorder scattering, the usage of EPs is preferred as well. This can be clearly seen in Fig.~\ref{fig:EPTrans} (c, d), where the anisotropy of the signal survives even for disorders with correlation amplitude twice larger compared to exciton (no transport anisotropy was observed for exciton transport at $\sigma=0.3\,\mathrm{meV}$). Therefore, this clearly demonstrates, that despite the fact, that considerably positively detuned EPs should be used in the experiment to observe transport anisotropy, the usage of EPs is preferred compared to bare exciton system.

\section{Conclusions}
\label{Conclusions}
In conclusion we have presented a method to experimentally observe the peculiar structure of exciton dispersion in TMDs. In particular, we analyzed transport properties of excitons and EPs, when the system is pumped close to the bottom of the quasiparticle band. Pumping with linearly polarized pump resulted in anisotropic photoemission compared to circularly polarized pump. We showed that the effect is related to the large group velocity difference between two branches of quasiparticles due to the linear component of dispersion for one of the branches. We have analyzed the effect of disorder on the observed results and showed that the results persist for the moderate levels of disorder in the system. Finally, we demonstrated that similar effect should be present for positively detuned EPs as well. The usage of EPs in this regard is preferred over excitons, despite the fact that in considerably positively detuned EPs the photon proportion is small. Our results shows that upon the development of better quality samples, exciton and EP transport can be used to develop optical switches which does not require the realization of superfluid phases. Given the experimental observation of quantum coherent EPs in TMDs, even at room temperature, applications of EPs in these materials for development of quantum switches is particularly interesting and deserves further studies.

\section{Acknowledgment}                                                 
We acknowledge fruitful conversations with Vinod Menon. This research was supported under National Science Foundation Grants EFRI-1542863, CNS-0958379, CNS-0855217, ACI-1126113 and the City University of New York High Performance Computing Center at the College of Staten Island.
\bibliography{bibexciton}
\bibliographystyle{apsrev4-1}

\end{document}